\documentclass[runningheads]{llncs}

\usepackage{graphicx}
\usepackage{amsmath}
\usepackage{amsbsy}
\usepackage{changepage}
\usepackage{float}
\usepackage{amsfonts}
\usepackage{wrapfig}
\usepackage{hyperref}
\usepackage{xcolor}

\begin{document}

\title{Attenuation Imaging with Pulse-Echo Ultrasound based on an Acoustic Reflector}

\titlerunning{Pulse-Echo Attenuation Imaging}

\author{Richard Rau \and Ozan Unal \and Dieter Schweizer \and \\Valery Vishnevskiy \and Orcun Goksel}
\authorrunning{R. Rau et al.}

\institute{Computer-assisted Applications in Medicine, ETH Zurich, Switzerland}

\maketitle          

\begin{abstract}
Ultrasound attenuation is caused by absorption and scattering in tissue and is thus a function of tissue composition, hence its imaging offers great potential for screening and differential diagnosis. 
In this paper we propose a novel method that allows to reconstruct spatial attenuation distribution in tissue based on computed tomography, using reflections from a passive acoustic reflector. 
This requires a standard ultrasound transducer operating in pulse-echo mode, thus it can be implemented on conventional ultrasound systems with minor modifications.
We use calibration with water measurements in order to normalize measurements for quantitative imaging of attenuation.
In contrast to earlier techniques, we herein show that attenuation reconstructions are possible without any geometric prior on the inclusion location or shape.
We present a quantitative evaluation of reconstructions based on simulations, gelatin phantoms, and \textit{ex-vivo} bovine skeletal muscle tissue, achieving contrast-to-noise ratio of up to 2.3 for an inclusion in \textit{ex-vivo} tissue.
\keywords{ultrasound  \and attenuation \and computed tomography \and speed of sound  \and limited angle tomography}
\end{abstract}

\section{Introduction}
Changes in tissue characteristics may be a prominent indication of pathology, which can be probed by sonography.
For instance, shear-wave elastography aims to estimate tissue shear-modulus~\cite{sandrin_shear_2002,eby_validation_2013}, while speed-of-sound imaging relates to tissue bulk modulus~\cite{duric_detection_2007,sanabria_spatial_2018}.

Typical B-mode ultrasound images the amplitude of echos from tissue.
The ultrasound (US) intensity \emph{attenuates} during acoustic propagation via several mechanisms: US waves may \emph{reflect} and \emph{scatter}, respectively, from large and small tissue structures of differing acoustic impedance.
Frictious losses in tissue cause \emph{viscous absorption}.
Additionally, a main mode of energy loss in tissue is via \emph{relaxation absorption}, which is due to consecutive wave-fronts ``hitting'' the tissue that is locally recovering (bouncing back) from the push of an earlier wave-front~\cite{smith_introduction_2011}. 
Overall, the effects above lead to \emph{ultrasound attenuation} (UA), i.e. the amplitude decay of US signals, dependent on tissue composition; e.g.\ UA is known to differ between malignant and benign tissues such as in breast tumors~\cite{goss_comprehensive_1978,goss_compilation_1980,bamber_ultrasonic_1979,bamber_acoustic_1981}.
Therefore, imaging of UA can serve as a diagnostic bio-marker.

Successful imaging of UA has so far only been achieved using complex, dedicated imaging setups using transmission mode, e.g.\ a ring transducer scanning the breast suspended in a water bath~\cite{duric_detection_2007,li_breast_2017}. 
Such transmission mode setups cannot be implemented with conventional clinical US systems with hand-held transducers, making UA imaging inaccessible for most clinical practice. 
In this paper we propose a novel method for imaging spatial UA distribution based on \emph{limited-angle computed tomography} (LA-CT) with a conventional linear array transducer. 
The only additional hardware required is a passive reflector, similarly to those proposed for speed-of-sound imaging in~\cite{sanabria_speed--sound_2018}.
A reflector setup was also proposed earlier for quantifying UA~\cite{huang_ultrasonic_2005,chang_reconstruction_2007}, which was however not suitable for imaging (reconstructing) arbitrary spatial distributions, but shown only for quantifying values for known geometries.
Furthermore, due to reconstruction instabilities, only synthetic and phantom examples could be quantified, but no actual tissue samples.
In this work we present for the first time the image reconstruction of acoustic attenuation in tissue using a single, conventional ultrasound transducer.

\section{Methods}
\begin{figure}[t]
\includegraphics[width=\textwidth]{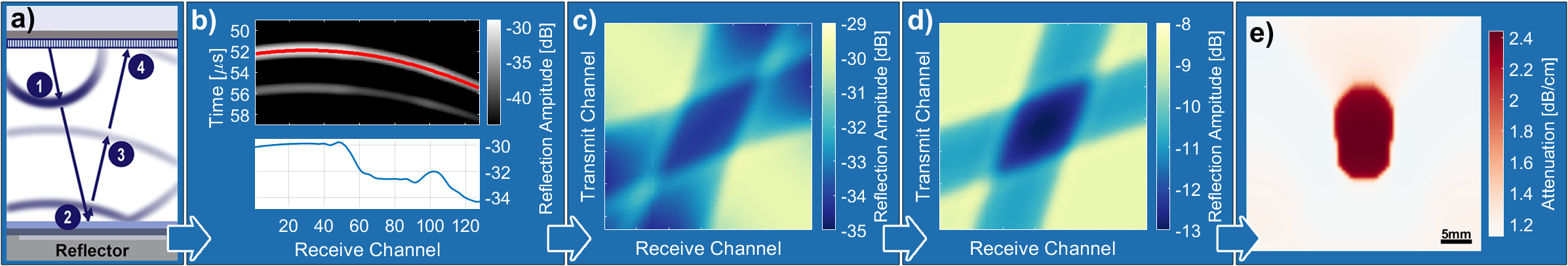}
\caption{Processing chain for the attenuation reconstruction with the reflector setup. \textbf{a)} Schematic of the passive reflector setup with exemplary wavefronts. \textbf{b)} k-Wave simulated received echos from the passive reflector. \textbf{c/d)} Reflected amplitudes for all $M=128^2$ transmit/receive channel combinations before (c) and after calibration (d). \textbf{e)} Reconstruction of the UA distribution.
} \label{fig:proc}
\end{figure}

In Fig.~\ref{fig:proc} an overview of the acquisition and processing chain for the UA reconstruction is illustrated. A plexiglas plate (density: $\rho$=$1180\,k\mathrm{g/m^3}$, speed-of-sound: $c$=$2700\,\mathrm{m/s}$) is placed at a distance $d$ away from the transducer.
We employ a full-matrix (multistatic) acquisition sequence, where following each single element transmit (Tx), echo on all elements is received (Rx) in parallel. Such process is then repeated for the transmission of all channels.
A sample wavefront path is shown in Fig.~\ref{fig:proc}a at different time-points: after transmission (1), while being reflected from the plate (2), during echo travel (3), and during reception at the transducer (4). 
The echo from the passive reflector's top surface is then delineated across Rx channels for each transmit event and the amplitude of the signal envelope along this delineated reflection is recorded as seen in Fig.~\ref{fig:proc}b. 
Reflection amplitude for all Tx-Rx combinations are shown in Fig.~\ref{fig:proc}c. 
With the approximation that the ultrasound pulses propagate as rays, the amplitude of the reflector when transmitting with channel $t$ and receiving with channel $r$ is described by 
\begin{equation}  \label{eq:A}
    A_{t,r} = A_{t,\theta}\cdot R(\theta) \cdot S_{r} \cdot \exp\bigg(\!\!-\!\!\!\int_{\text{ray}_{t,r}} \!\!\!\! \alpha(x,y)\ \mathrm{d} l \bigg),
\end{equation}
where $S_{r}$ is the sensitivity of Rx element $r$, $R(\theta)$ is the incident-angle dependent reflection coefficient at the reflector interface, $A_{t,\theta}$ is the initial amplitude of channel $t$ in ray-direction $\theta$, and the exponent describes the amplitude decay based on the line integral of attenuation $\alpha$ along ray$_{t,r}$ from element $t$ to $r$. 

\subsection{Calibration}
In order to isolate the attenuation effects from $A_{t,r}$, one needs to estimate or compensate for any other influences in~(\ref{eq:A}) such as from the impulse response of the transducer (affecting $A_{0,t}$ and $ S_{r}$) and reflection characteristics (i.e., $R(\theta)$).
To that end, we normalize the measurements with a calibration experiment in water, for which the speed-of-sound $c_{\mathrm{water}}= 1482.5$\,m/s and the attenuation coefficient $\alpha_{\mathrm{water}} \approx 0.05$\,Np/cm are known from the literature, given water temperature $T$$=$$20^{\circ}$\,C and imaging frequency of 5\,MHz.
For the calibration experiment and an actual acquisition, $A_{t,\theta}$ and $S_{r}$ can be assumed to be similar; however, $R(\theta)$ may differ due to a speed-of-sound mismatch between water and tissue.
Nevertheless, such reflection coefficient at the acoustic reflector interface can be analytically estimated using Snell's law, because the wavelength is smaller compared to the reflector dimensions.
For the reflector interface:
\begin{equation}
    R_{k}(\theta) = \frac{m_{k}\cos(\theta)-n_{k}\sqrt{1-\frac{\sin^2(\theta)}{n_{k}^2}}}{m_{k}\cos(\theta)+n_{k}\sqrt{1-\frac{\sin^2(\theta)}{n_{k}^2}}},
\end{equation}
where speed-of-sound ratio $n_{k}$=$c_{\mathrm{reflector}}/c_{k}$ and density ratio $m_{k}$=$\rho_{\mathrm{reflector}}/\rho_{k}$.
We herein assume $\rho_{\mathrm{tissue}}$$\approx$$\rho_{\mathrm{water}}$$\approx$1000\,kg/m$^3$. 
The normalized reflection amplitude matrix (cf.\ Fig.\,\ref{fig:proc}d) can therefore be computed  as:
\begin{align}\label{eq:normA}
    b_{t,r} = \ln \frac{A_{t,r,\mathrm{tissue}}(\theta) R_{\mathrm{water}}(\theta)}{A_{t,r,\mathrm{water}}(\theta)R_{\mathrm{tissue}}(\theta)}  = -\int_{\text{ray}_{t,r}} \!\!\!\! \alpha\ \mathrm{d}l  \approx -\!\!\!\sum_{i\in\text{ray}_{t,r}}\!\!l_{i}\alpha_{i}\,,
\end{align}
where the ray integral is discretized as summation over a reconstruction grid, with each attenuation value $\alpha_i$ weighted by path length $l_i$ within grid element $i$.

\subsection{Attenuation Reconstruction}
\begin{figure}[t]
\centering
    \includegraphics[width=.42\textwidth]{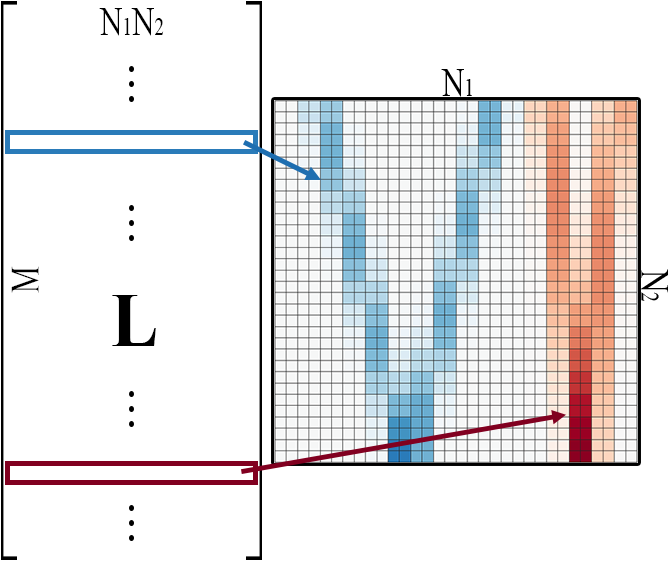}
    \caption{Reconstruction matrix $\textbf{L}$ with two representative paths depicted.} \label{fig:reconst}
\end{figure}

Given $M$ logarithms of normalized reflection amplitudes $\mathbf{b}\in\mathbb{R}^{M}$, we perform a tomographic reconstruction of the spatial UA distribution on a  $N_1$$\times$$N_2$ spatial grid by formulating the following convex optimization problem:\hfill
\begin{equation}\label{eq:opti}
 \boldsymbol{\hat\alpha} = \underset{\boldsymbol{\alpha}}{\arg\min}
 \| \textbf{L}\boldsymbol{\alpha}  + \mathbf{b} \|_1  +  \lambda \|\textbf{D}\boldsymbol{\alpha} \|_1,
\end{equation}
where $\textbf{L}\in\mathbb{R}^{M\times N_1N_2}$ is the sparse ray path matrix (cf.\ Fig.\,\ref{fig:reconst}) that implements~(\ref{eq:normA}) and $\boldsymbol{\alpha}\in\mathbb{R}^{N_1N_2}$ is the reconstructed image.
A regularization weight $\lambda$ controls the amount of spatial smoothness and is essential due to the ill-conditioning of $\mathbf{L}$.
The regularization matrix $\mathbf{D}$ implements LA-CT specific image filtering aimed to suppress streaking artifacts along wave propagation directions via anisotropic weighting of horizontal, vertical and diagonal gradients as described in~\cite{sanabria_hand-held_2016}. 
In this paper we empirically set $\lambda$=0.6 for all experiments, and use an unconstrained optimization package \texttt{minFunc}\footnote{\url{https://www.cs.ubc.ca/~schmidtm/Software/minFunc.html}} to numerically solve~(\ref{eq:opti}).

\section{Experiments}
\noindent {\bf Metrics. } We used the following metrics for quantitative analysis:
\vspace{-4pt}
\begin{itemize}
    \item Contrast-ratio fraction: $\mathrm{CRF} =  \hat C/C^*$, where $C = 2|\mu_{\mathrm{inc}} - \mu_{\mathrm{bkg}}|/(|\mu_{\mathrm{inc}}|+|\mu_{\mathrm{bkg}}|)$ with the mean inclusion value $\mu_{\mathrm{inc}}$ and mean background value $\mu_{\mathrm{bkg}}$. 
    Th$e\,\,\hat{}\,$  and $^*$ indicate the reconstruction and ground truth, respectively. 
    \item Contrast-to-noise ratio: $\mathrm{CNR} = |\mu_{\mathrm{inc}}-\mu_{\mathrm{bkg}}|/\sqrt{\sigma_{\mathrm{inc}}^2 + \sigma_{\mathrm{bkg}}^2}$,   variance $\sigma^2$. 
    \item Root-mean-squared-error: 
    $\mathrm{RMSE} =\sqrt{\|\hat{\boldsymbol{\alpha}}-\boldsymbol{\alpha}^\star\|_2^2 / N}$.
    \item Peak signal-to-noise ratio: $\mathrm{PSNR} = 20\log_{10}(\hat\alpha_\mathrm{max}/\mathrm{RMSE})$.
\end{itemize}
\vspace{-4pt}
Note that only $\mathrm{CNR}$ above can be computed without a given ground truth UA. 

\begin{figure}[t]
\includegraphics[width=\textwidth]{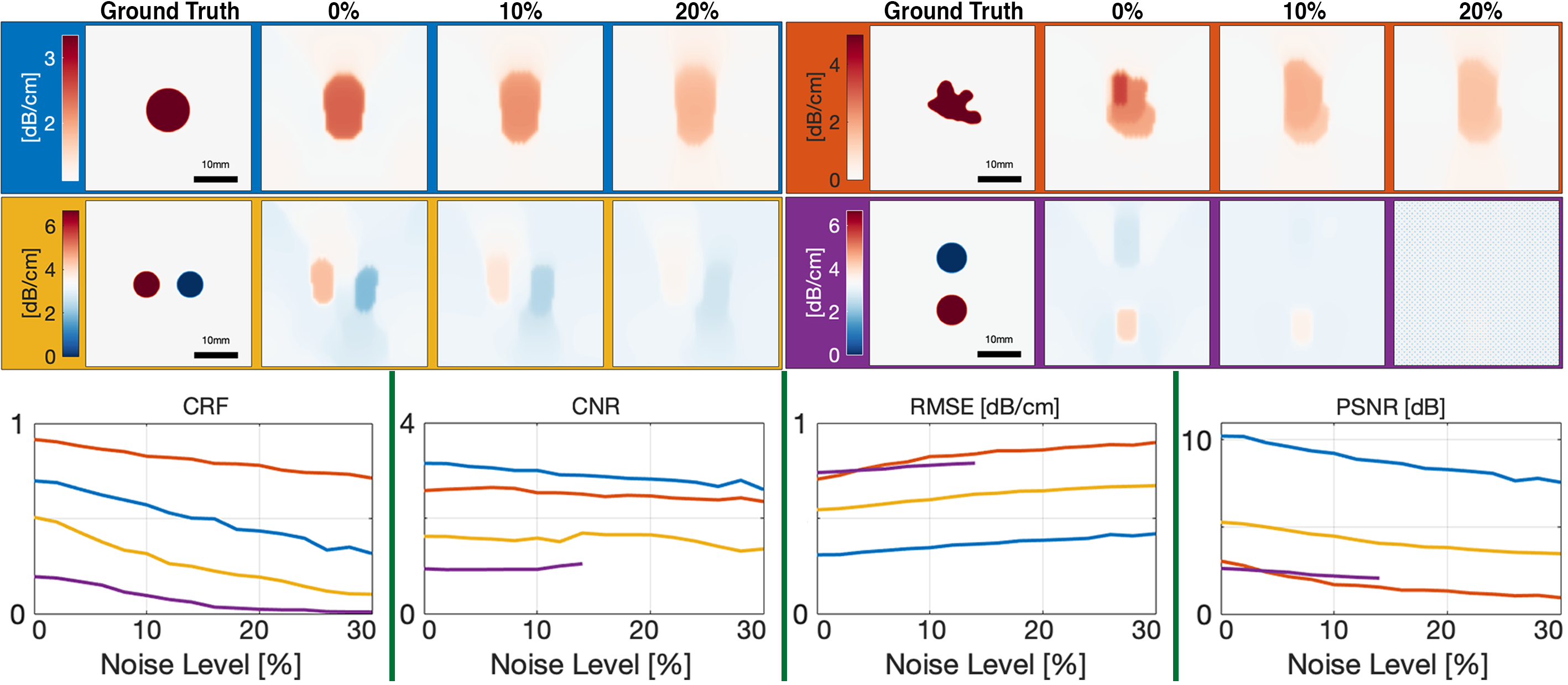}
\caption{Evaluation results for k-Wave simulated datasets at different noise settings. Box colors on the top correspond to the colors in the evaluation plots at the bottom. The purple case is only plotted up to 13\% noise, due to reconstruction failure at a higher noise level. The scale-bars represent $10\,\mathrm{mm}$.
} \label{fig:sim}
\end{figure}

\vspace{1ex} \noindent {\bf Simulation Study. }
To evaluate and quantify the accuracy of the UA reconstructions, four different simulations with increasing complexity in the UA patterns were performed, which are shown in Fig.\ref{fig:sim}.  
Simulations were performed using the k-Wave ultrasound simulation toolbox~\cite{treeby_k-wave:_2010} using a spatial grid resolution of $37.5\,\mu \mathrm m$.
Full-matrix acquisition was simulated at a center frequency of $5\,\mathrm{MHz}$ with pulses of $10$ half cycles, where longer pulse lengths allow for narrower bandwidth and more accurate estimation of the reflection amplitude based on the envelope at reflector delineation.
The transducer was simulated containing 128 channels (yielding $M = 128^2$) with a pitch of $300\,\mu \mathrm m$.
To investigate the effect of noise on reconstructions, we added zero-mean Gaussian noise on simulated measurements, with a standard deviation as a percentage of the normalized reflection amplitude matrix, an example of which is illustrated in Fig.\,\ref{fig:proc}d. 

\vspace{1ex} \noindent {\bf \textit{Ex-vivo} and Phantom Study. }
Gelatin phantoms were prepared with 10\% gelatin in water per weight. 
We created pure and scattering phantoms; the latter with 1\% Sigmacell Cellulose Type 50 (Sigma Aldrich, St. Louis, MO, USA). 
Fresh bovine skeletal muscle was used as an \textit{ex-vivo} tissue sample.
Using combinations of the above, we created four different phantoms: \textit{ex-vivo} muscle inclusions in gelatin phantom (a) with and (b) without cellulose for scattering; and \textit{ex-vivo} muscle tissue with embedded gelatin inclusions (c) with and (d) without cellulose, as shown in Fig.\ref{fig:exvivo}(left).

Measurements were carried out with the samples submerged in distilled water at room temperature, with muscle fibers oriented orthogonal to the imaging plane. 
This was to ensure that the acoustic wave propagation was always perpendicular to fiber direction, in order to avoid direction-dependent speed-of-sound and UA variation.
For the data acquisition we used a Verasonics Vantage 128 channel system connected to a Philips L7-4 transducer (Verasonics, Kirkland, WA, USA). 
Analogous to the simulation setup, we used a Tx center frequency of $5\,\mathrm{MHz}$ and a pulse length of $10$ half cycles, which was empirically found to be best suited for amplitude detection of reflections.
Reflector distance varied from 30\,mm to 46\,mm depending on the sample size. 
For normalization of the acquired \textit{ex-vivo}/phantom amplitude matrices, we conducted calibration measurements at the given reflector depths, in distilled water at room temperature. 

\section{Results and Discussion}

The four representative cases of the simulation study in Fig.\ref{fig:sim} demonstrate that our method accurately reconstructs the background UA values, the inclusion locations, and their approximate shape. 
Due to the regularization, the inclusion UA values are slightly underestimated w.r.t. the background, which is also reflected in CRF values being $<$1. 
At higher noise levels as well as with increasing complexity of the ground truth UA distribution, the accuracy of the reconstruction decreases, yielding higher RMSE and lower PSNR and CNR as one might expect. 
A generally observed feature is that the reconstructed inclusion shapes are axially elongated, which is a known problem for LA-CT reconstructions~\cite{sanabria_hand-held_2016}, especially at relatively higher noise levels. 
It can also be observed that the laterally separated inclusions are reconstructed more accurately than axially separated ones (Fig.\,\ref{fig:sim}). 
This demonstrates a general limitation of the LA-CT imaging: decreased resolution in axial direction due to insufficient spatial encoding in the transverse plane.
Still, it can be seen that even for this very challenging case, an approximate reconstruction of inclusion locations and attenuation characteristics is possible, at least at relatively lower noise levels.

\begin{figure}[t]
\includegraphics[width=\textwidth]{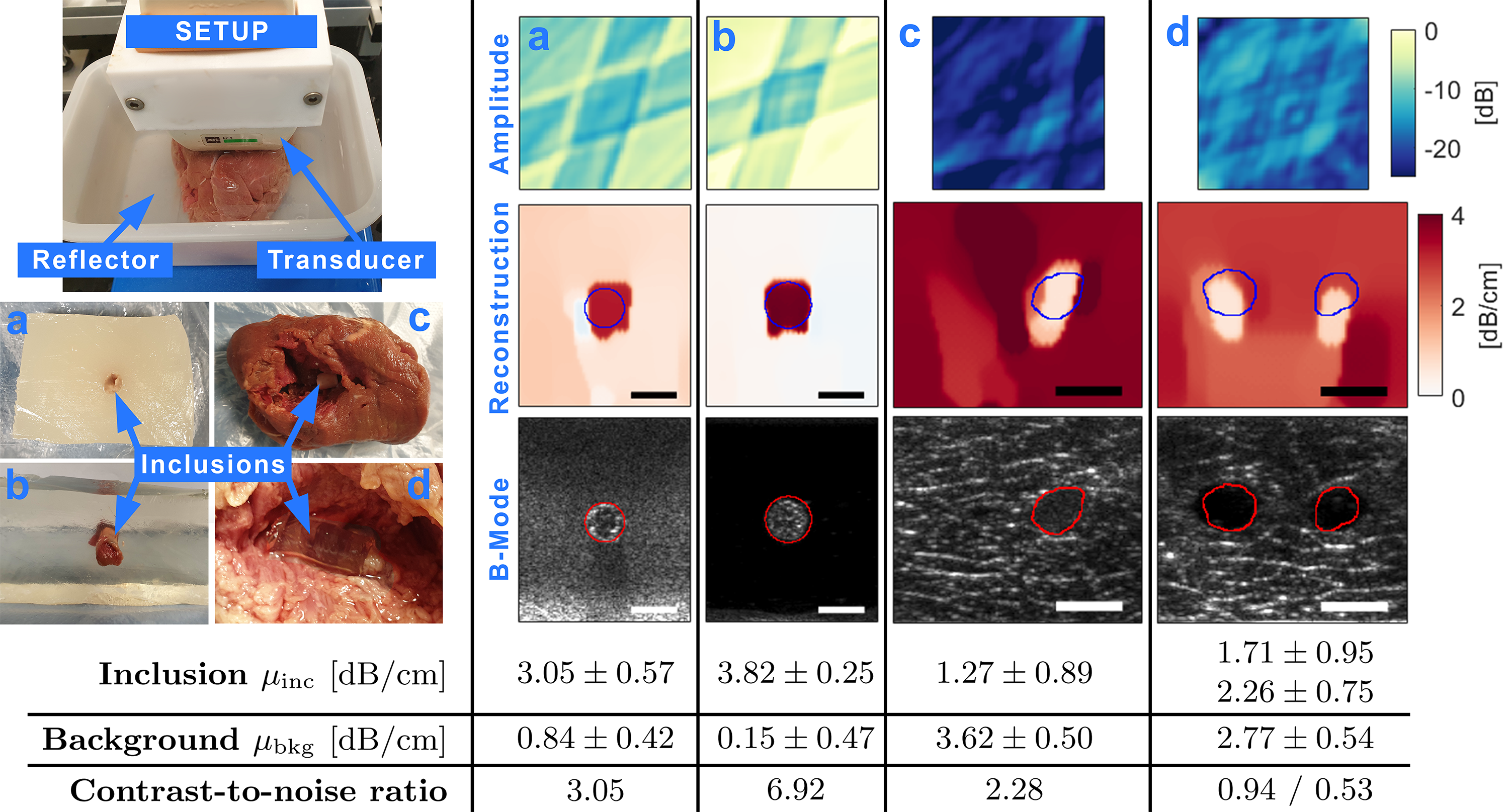}
\caption{\textit{Ex-vivo} bovine skeletal muscle results. On the left, the experimental setup is shown with the different study cases: muscle inclusion in gelatin (a) with and (b) without cellulose; and muscle samples with gelatin inclusions (c) with and (d) without cellulose. 
The scale-bars represent $10\,\mathrm{mm}$.}
\label{fig:exvivo}
\end{figure}

For the \textit{ex-vivo}/phantom experiments the results are shown in Fig.\,\ref{fig:exvivo} with the  normalized amplitude matrices, UA reconstructions and B-Mode images. 
For the pure gelatin phantom~(b), UA values are expected to be very low, which is corroborated with our finding of $\mu_\mathrm{bkg}=(0.15\pm.47)$\,dB/cm as the mean background attenuation value.
UA is expected to increase with added cellulose and hence scattering, which again is confirmed by our finding of $\mu_\mathrm{bkg}=(0.84\pm 0.42)\,\mathrm{dB/cm}$ for the phantom~(a).
Furthermore, the tissue inclusions in (a\&b) are reconstructed successfully, with high CNR values of 3.05~(a) and 6.92~(b), when the inclusion is delineated using the B-Mode images.

For the cases where the gelatin inclusions are placed in the muscle tissue, the reconstructions also perform well, even though the amplitude matrices show no clear and distinct profile, as was observed above for the cases (a\&b).
The inclusions in UA reconstructions are observed to be axially elongated, similarly to the simulated cases. 
This smearing results in an underestimation of UA inclusion values in the delineated regions compared to the background values, thus leading to a decrease in CNR.
To verify the robustness of our method for the \textit{ex-vivo} experiments, muscle UA values across all four experiments (a-d) are compared in the table in Fig.\,\ref{fig:exvivo}.
All muscle values (which are inclusion in a and b, and background in c and d) are seen to lie within $(3.21\pm0.67)$\,dB/cm, which is in agreement with the values reported in the literature for bovine skeletal muscle when measured perpendicular to fibers~\cite{goss_comprehensive_1978}.

Note that local speed-of-sound variations cause US wavefront aberrations; for instance for inclusions, an acoustic lens effect is observed where the amplitude readings from straight ray approximation are inaccurate.
This is visible in Fig.\,\ref{fig:exvivo}a in the amplitude matrix, where higher amplitudes are observed right on the margin of the relatively higher attenuating (darker) cross pattern. 
To improve the reconstructions further, a possibility would be to correct ray refractions based on speed-of-sound estimations, which can be derived from timing deviations~\cite{sanabria_speed--sound_2018}. 

In this study we only compensate for the speed-of-sound effects on the reflection coefficient differences at the reflector interface, based on~(\ref{eq:normA}). 
However, tissue speed-of-sound variations may additionally be affecting the angular beam profile of Tx/Rx transducer elements, thus introducing deviations in the amplitude matrix hindering its effective normalization. 
Nevertheless, given the successful reconstruction results, we believe such effect on the beam profile to be minimal.

UA is dependent on the frequency of the US pulse, i.e.\ $\propto f^y$, where $y$ is tissue dependent. 
Thus, using narrowband pulses, reconstructions could be carried out at different frequencies to estimate the frequency dependence parameter $y$ as yet another imaging biomarker and tissue characteristic. 

For multi-parametric characterization, UA can be an imaging biomarker complementary to speed-of-sound, which may be superior to elastography~\cite{glozman_method_2010}. The quantification of speed-of-sound was recently proposed for breast density~\cite{Sanabria_breast-density_18} and sarcopenia assessment~\cite{Sanabria_speed_18}.

A practical limitation of our proposed UA imaging method is that an algebraic reconstruction is utilized.
A variational network solution similar to~\cite{vishnevskiy_deep_2019,vishnevskiy2018image} with inference times on the order of milliseconds could help to overcome this limitation towards real-time UA imaging.

\section{Conclusion and Outlook}
In this paper we have presented a novel approach for reconstructing ultrasound attenuation distribution in tissue, known to be relevant as imaging biomarker, e.g., for differentiating malignant tumor structures.
We evaluated sensitivity w.r.t. noise and domain complexity with simulations and \textit{ex-vivo} experiments.
Inclusion size or shape may not always be known in many clinical applications. 
We show herein, to best of our knowledge for the first time, accurate reconstruction of attenuation without prior knowledge using conventional ultrasound linear arrays.
Since our proposed method can be implemented in standard ultrasound systems and requires only minimal hardware addition of a passive acoustic reflector, it is readily translatable to clinical setting. 
Prospective applications could be the anatomical locations that allow two-sided access, such as the imaging of the breast and the extremities.

\vspace{5ex} \noindent {\bf Funding} was provided by the Swiss National Science Foundation and Innosuisse.

\bibliographystyle{splncs04}
\bibliography{references_selected}

\end{document}